\documentclass[aps,prl,floatfix,twocolumn,superscriptaddress]{revtex4-1}
\usepackage{amssymb}
\usepackage{amsmath}
\usepackage{graphicx}
\usepackage{color}

\begin{document}

\title{Electron quantum optics : partitioning electrons one by one}

\author{E. Bocquillon}

\author{F.D. Parmentier}
\affiliation{Laboratoire Pierre Aigrain, Ecole Normale Sup\'erieure, CNRS (UMR 8551), Universit\'e P. et M. Curie, Universit\'e D. Diderot, 24 rue Lhomond, 75231 Paris Cedex 05, France.}

\author{C. Grenier}
\affiliation{Universit\'e de Lyon, F\'ed\'eration de Physique Andr\'e Marie Amp\`ere, CNRS - Laboratoire de Physique de l'Ecole Normale Sup\'erieure de Lyon, 46 All\'ee d'Italie, 69364 Lyon Cedex 07,France.}

\author{J.-M. Berroir}
\affiliation{Laboratoire Pierre Aigrain, Ecole Normale Sup\'erieure, CNRS (UMR 8551), Universit\'e P. et M. Curie, Universit\'e D. Diderot, 24 rue Lhomond, 75231 Paris Cedex 05, France.}

\author{P. Degiovanni}
\affiliation{Universit\'e de Lyon, F\'ed\'eration de Physique Andr\'e Marie Amp\`ere,
CNRS - Laboratoire de Physique de l'Ecole Normale Sup\'erieure de Lyon,
46 All\'ee d'Italie, 69364 Lyon Cedex 07,France.}

\author{D.C. Glattli}

\author{B. Pla\c cais}
\affiliation{Laboratoire Pierre Aigrain, Ecole Normale Sup\'erieure, CNRS (UMR 8551), Universit\'e P. et M. Curie, Universit\'e D. Diderot, 24 rue Lhomond, 75231 Paris Cedex 05, France.}

\author{A. Cavanna}

\author{Y. Jin}
\affiliation{CNRS, Laboratoire de Photonique et de Nanostructures (LPN), Route de Nozay, 91460 Marcoussis, France}

\author{G. F\`eve }
\email{feve@lpa.ens.fr}
\affiliation{Laboratoire Pierre Aigrain, Ecole Normale Sup\'erieure, CNRS (UMR 8551), Universit\'e P. et M. Curie, Universit\'e D. Diderot, 24 rue Lhomond, 75231 Paris Cedex 05, France.}

\date{\today}

\begin{abstract}
We have realized a quantum optics like Hanbury Brown and Twiss (HBT) experiment by partitioning, on an electronic beam-splitter, single elementary electronic excitations produced one by one by an on-demand emitter. We show that the measurement of the output currents correlations in the HBT geometry provides a direct counting, at the single charge level, of the elementary excitations (electron/hole pairs) generated by the emitter at each cycle. We observe the antibunching of low energy excitations emitted by the source with thermal excitations of the Fermi sea already present in the input leads of the splitter, which suppresses their contribution to the partition noise. This effect is used to probe the energy distribution of the emitted wave-packets.
\end{abstract}

\maketitle

The development of quantum electronics based on the coherent manipulation of single to few quasi-particles in a ballistic quantum conductor has raised a strong interest in the recent years \cite{Keeling2006,Ol'khovskaya08, Splettstoesser2009, Grenier2011,Bertoni2000,Feve2007, Blum2007}. On the theoretical side,  many proposals have suggested to generate and manipulate single electronic excitations in optics like setups \cite{Ol'khovskaya08, Splettstoesser2009, Grenier2011} and to use them in Fermion based quantum information processing \cite{Bertoni2000}. On the experimental side,  triggered electron sources that supply single electron states on-demand have been demonstrated \cite{Feve2007, Blum2007} but there has been no report so far of their implementation in an electron quantum optics experiment (i.e electron optics at the single charge level). Actually, the very principle of electron quantum optics is still under question as singling out a single elementary excitation remains a complex issue \cite{Degio2009} in solid state where the Fermi sea builds up from many interacting electrons.

In this work, we have realized the partitioning of single electron/hole excitations emitted one by one by the on demand electron source we recently developed \cite{Feve2007} using an electronic beam splitter in the Hanbury Brown and Twiss geometry \cite{HBTelec}. From low frequency current correlations measurements, we count the number of elementary excitations produced by the source at the single charge level. We also demonstrate that the random partitioning of low energy excitations produced by the source is suppressed by their antibunching with thermal excitations of the Fermi sea. This quantum effect provides an efficient tool to probe the energy distribution of the individual quantum states produced by the source. By tuning the emission parameters we show that the energy distribution can be shaped in a controlled manner. Finally, this work defines the proper conditions for the manipulation of a single elementary excitation in the presence of a thermal bath.

Electron quantum optics, like its photonic counterpart, relies on the  manipulation of single particle states supplied on-demand and characterized by the measurements of current-current correlations. The study of current-current correlations in quantum conductors has been widely used to probe the statistics of particles emitted by a source. The most common source is the  DC biased contact which
produces a stationary current where the electronic populations are those of a degenerate Fermi gas at thermal equilibrium. The Pauli exclusion
principle then enforces a noiseless flow of electrons \cite{noiseth} which has been probed through auto-correlation measurements \cite{noiseexp} or cross-correlation in the HBT geometry \cite{HBTelec}. Despite its ability to naturally produce noiseless single electron beams, such a continuous source cannot produce electron states with a defined timing. The controlled manipulation of single electrons requires to replace stationary (DC driven) by triggered (AC driven) single particle emitters.

AC sources differ from DC sources as their elementary emission processes consist in the generation of coherent electron/hole pairs \cite{Vanevic2008, Keeling2008}, so that the electron and hole populations deviate from equilibrium. As a first consequence, contrary to DC sources, no information can be gained from low frequency noise measurement of the current directly emitted by the source, as there is no charge transfer or charge fluctuations on long times, the electron and hole currents compensating each other. The statistics of charge transfer is then revealed in the high frequency noise \cite{Mahe2010}. However, the number of elementary excitations produced by the source is hardly extracted from such measurements since an electron/hole pair is detected only if the delay between the two particles is larger than the temporal resolution of the setup. It is known that low frequency noise can be recovered from the random and independent partitioning of electrons and holes on an electronic beam-splitter \cite{noiseac, Reydellet03}. The information one can extract from noise measurements in the HBT geometry then strongly depends on the nature of the source. While for a DC emitter \cite{HBTelec}, the low frequency correlations of the current can reveal the fluctuations of the number of particles (electrons) emitted by the source, when dealing with an AC emitter, the same measurement instead yields the average number of elementary excitations (electron/hole pairs) generated by the source at each of its cycles.

Considering a periodically driven emitter at frequency $f_d$ placed on input $1$ of a splitter independently transmitting electrons and holes with probability $T$, the average number of electron/hole pairs emitted in one period can be directly extracted from the low frequency correlations between the currents at the output $3$ and $4$,   $S_{3,4} = -2 e T(1-T) \langle I_{part} \rangle$, where $\langle I_{part} \rangle =  e f_d \langle N_{e}+N_{h} \rangle $ is the particle current, $\langle N_{e} \rangle$ and $\langle N_{h} \rangle$
being the average numbers of electrons and holes emitted per period (see supplementary material).

\begin{figure}[h]
   \centering\includegraphics[width=0.5\textwidth]{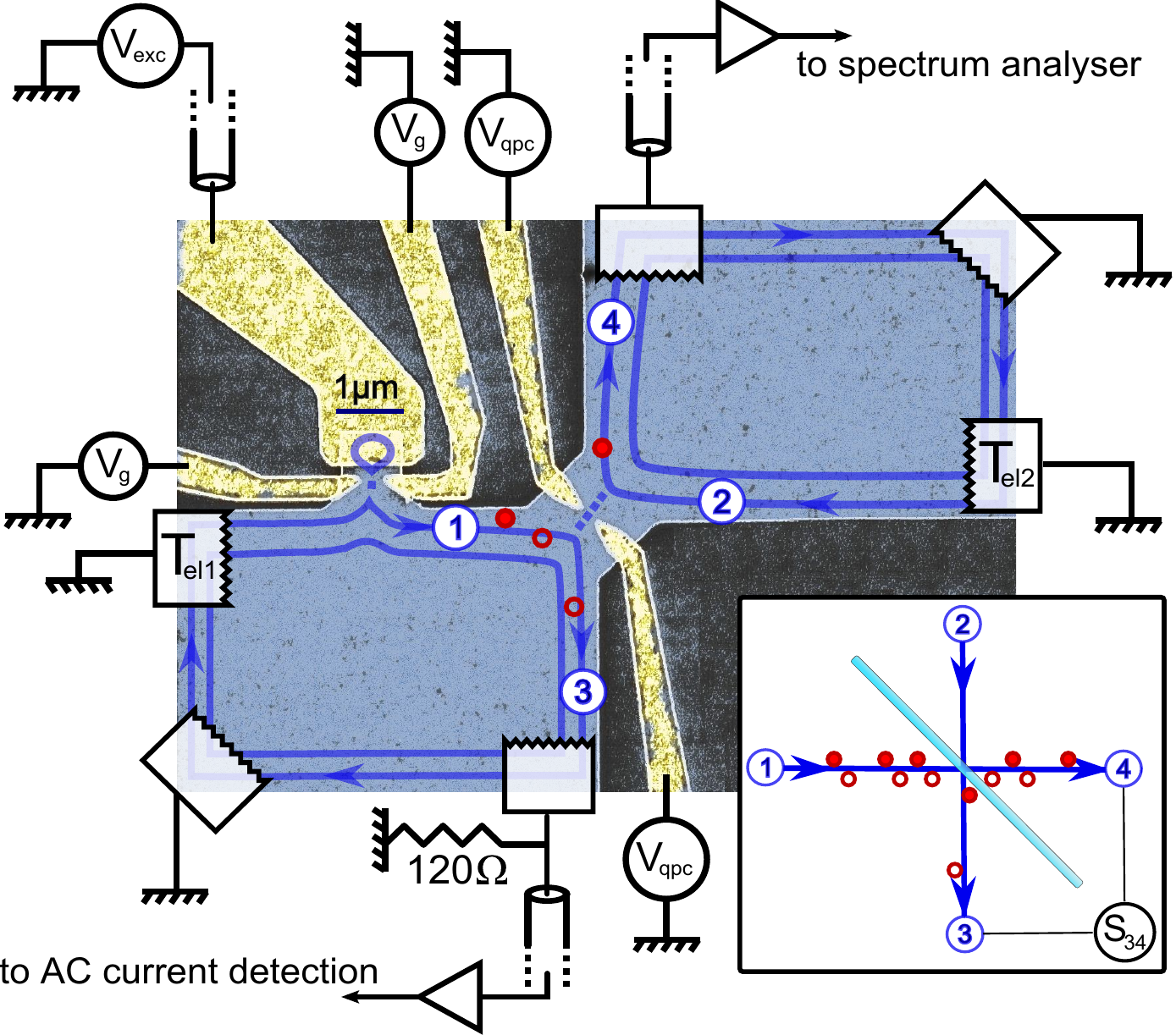}
  \caption{Hanbury Brown and Twiss experiment, sketch (inset) and sample. Schematic illustration based on the SEM picture of the sample. A perpendicular magnetic field $B=3.2$ T is applied in order to work at filling factor $\nu=2$. The two edge channels are represented by blue lines. The emitter is placed on input $1$, 2.5 microns before the electronic splitter whose gate voltage $V_{qpc}$ is set to fully reflect the inner edge while the outer edge can be partially transmitted
with tuneable transmission $T$.  The emitter is tunnel coupled to the outer edge channel with a transmission $D$ tuned by the gate voltage $V_g$. Electron emission is triggered by the excitation drive $V_{exc}(t)$. Average measurements of the AC current generated by the source are performed
on output $3$, whereas output $4$ is dedicated to the low frequency noise measurements $\delta S_{4,4}$. Inset: sketch of the Hanbury Brown and Twiss experiment. The average number of electrons (filled red dots) and holes (empty red dots) emitted on input $1$ can be extracted from the current correlations between outputs $3$ and $4$.   }
  \label{Fig1}
\end{figure}
However, there are deviations to this classical reasoning because the input arms are populated by thermal electron/hole excitations. These thermal excitations interfere with the ones produced by the source, affecting their partitioning.
Their antibunching with electron/hole excitations cannot be accounted by the classical description and one needs to rely on a quantum description (see Ref. \onlinecite{Grenier2011} or supplementary material). For clarity, let us first consider the effect of thermal excitations in input $2$, disregarding the effect of temperature in input $1$. This leads to :
\begin{eqnarray}
S_{3,4} & = &  T(1-T)\left[S_{2,2}-4 e^2 f_d \ N_{HBT}\right]\\
 N_{HBT} & = &  \frac{\langle N_{e} \rangle + \langle N_{h} \rangle }{2} - \int_{0}^{\infty} d\epsilon (n_e(\epsilon) + n_h(\epsilon)) f_2(\epsilon) \label{eq:dQ} \nonumber \\
 \\
\langle N_{e} \rangle & =& \int_{0}^{\infty} d\epsilon \ n_e(\epsilon) \\
\langle N_{h} \rangle & =& \int_{0}^{\infty} d\epsilon \ n_h(\epsilon) \label{eq:Nexc}
\end{eqnarray}
where $S_{2,2}$ is the low frequency thermal noise on input $2$ and $f_2(\epsilon)$ the equilibrium Fermi distribution at arm $2$ temperature $T_{el,2}$. The energy reference is the Fermi energy of the electron gas, i.e. $\epsilon_F=0$. $n_e (\epsilon)$ $\big($ respectively $n_h(\epsilon) \big)$ is the energy density of electronic (respectively hole) excitations added by the source during one period.
The HBT contribution $ N_{HBT}$ differs from the classical one $\frac{\langle N_e \rangle + \langle N_h \rangle }{2}$ by: $-\int_{0}^{\infty} d\epsilon (n_e(\epsilon) + n_h(\epsilon)) f_2(\epsilon)$. The minus sign reflects the antibunching of fermionic particles colliding on the splitter and replaces the plus sign observed for bosons, for example in the Hong Ou Mandel experiment \cite{HOM}. The number of detected electron/hole pairs $N_{HBT}$ is thus reduced by the energy overlap between the source excitations and the thermal ones. For a vanishing overlap, classical partitioning is recovered. For a non vanishing overlap, some of the source excitations cannot be distinguished from thermal
ones and do not contribute to the partition noise. This antibunching provides a powerful tool to probe the energy distributions of the excitations produced by the source.
In a real system, one should also take care of thermal excitations in arm $1$ emitted by the reservoir upstream of the source, which also interfere with the ones additionally produced by the source. For equal temperatures on both arms, $T_{el,1}=T_{el,2}$,
the cross-correlations $S_{3,4}$, or equivalently the excess auto-correlations $\delta S_{4, 4}$, directly measure the contribution of the excess excitations produced by the source :
\begin{eqnarray}
S_{3,4}  & = & -\delta S_{4,4} =   -4e^2 f_d \ T(1-T) \ \delta N_{HBT}\\
\delta N_{HBT} & = & \frac{ \langle \delta N_{e} \rangle + \langle \delta N_{h} \rangle }{2} - \int_{0}^{\infty} d\epsilon (\delta n_e(\epsilon) + \delta n_h(\epsilon)) f_2(\epsilon) \label{eq:dQ2} \nonumber \\
\end{eqnarray}
where $\delta$ refers to the difference between the on and off states of the source.

We now turn to the experimental realization of the HBT experiment using a single particle emitter. The quantum conductor is a two-dimensional electron gas in the quantum Hall regime.
Using one-dimensional chiral propagation along a quantum Hall edge channel, and a quantum point contact taken as an electronic beam-splitter, the geometry used in the seminal HBT experiment can be mimicked, as depicted on Fig.1. The emitter placed on input $1$ is a periodically driven mesoscopic capacitor \cite{capacitor}  made of a quantum dot (with level spacing $\Delta = 2.1$  K) tunnel coupled to input lead $1$ by a quantum point contact whose gate voltage $V_g$ tunes the dot/edge channel transmission $D$. A periodic RF drive, applied on a metallic top gate capacitively coupled to the dot, gives rise to the periodic emission of a single electron followed by a single hole. The top gate of the source is driven at frequency $f_d=1.7$ GHz using either a square wave (containing approximately three odd harmonics) or a sine wave, so as to engineer different single particle wave-packets.
As described in ref.\cite{Feve2007}, we adjust the emitter parameters so that the average charge $Q^t$, emitted from the dot in time $\frac{1}{2 f_d}$, equals the elementary charge $e$ for a large range of dot transmission $D$. For $D \approx 1$, $Q^t$ exceeds $e$ as the dot is fully open. $Q^t$ goes to zero for small $D$ as the average escape time $\tau$ becomes larger than the drive period. Finally $Q^t =e $ within $10$ percent for $0.2 \leq D \leq 0.7$).

\begin{figure}[h]
   \centering\includegraphics[width=0.5\textwidth]{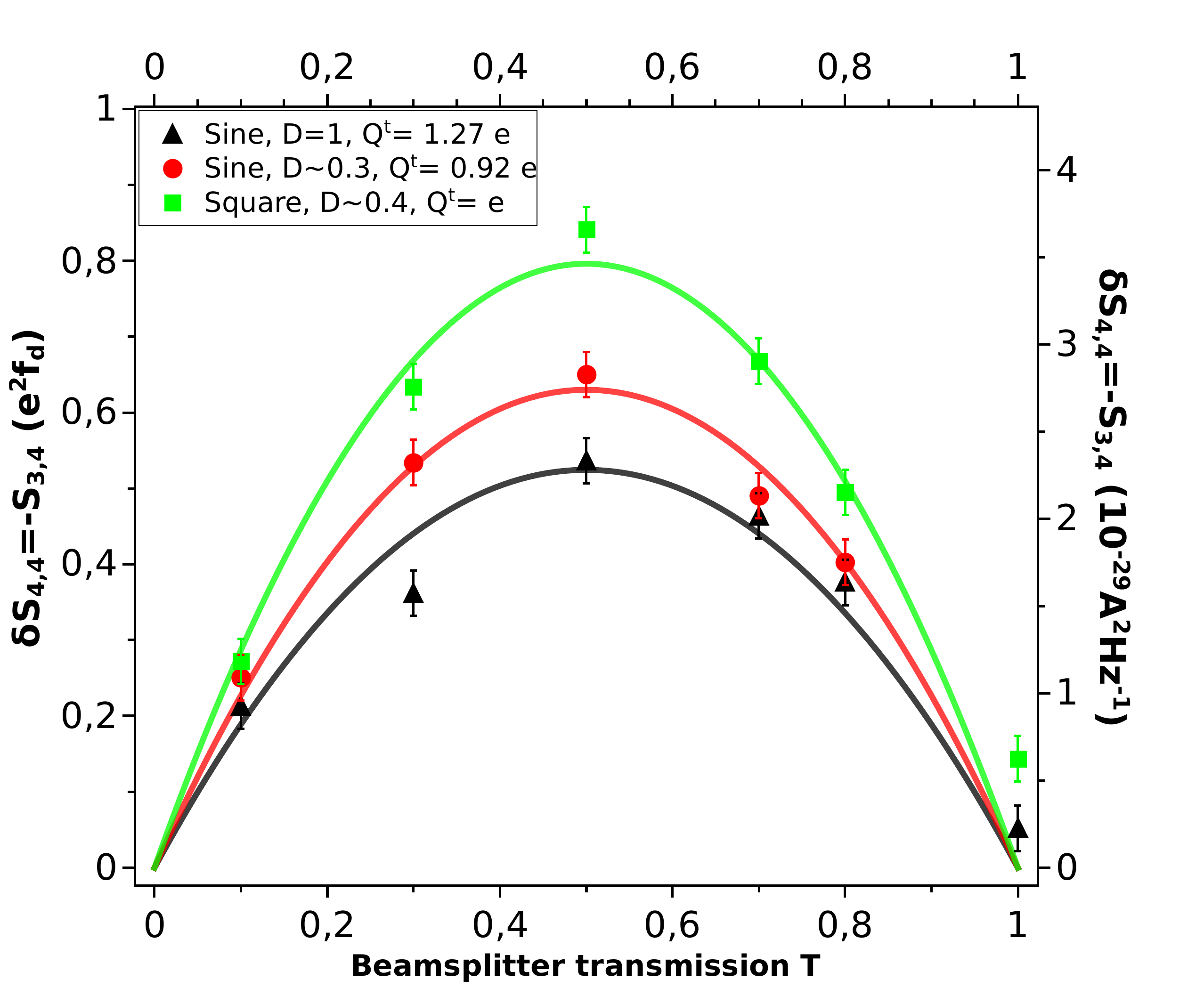}
  \caption{Low frequency correlations $\delta S_{4,4} = -S_{3,4}$ in units of $e^2 f_d$ (left axis) and $A^2.Hz^{-1}$ (right axis) as a function of the beam-splitter transmission $T$. Three types of RF drives are plotted, a sine wave at $D=1$ (black triangles), a sine wave at $D=0.3$ (red circles) and a square wave at $D=0.4$ (green squares). The solid lines represent adjustments with the expected $T(1-T)$ dependence.}
  \label{Fig2}
\end{figure}
Fig.2 presents measurements of the low frequency correlations $\delta S_{4,4}=-S_{3,4}$. The black and red dots are obtained using a sine wave drive at transmission
$D=1$ and $D=0.3$ while the green curve is obtained using a square wave for $D=0.4$. For all curves, the expected $T(1-T)$ dependence is observed, but the noise magnitudes (at $T=1/2$) notably differ and do not
reproduce the average transferred charge $Q^t$. For all three cases, $\delta N_{HBT}$ is
smaller than $1$, the value which should be observed for the classical partitioning of a single electron/hole pair. We attribute this discrepancy to the non-zero overlap between the source excitations and the thermal ones.  The highest value of $\delta N_{HBT}$ is observed for a square wave. In this case, a single energy level of the dot is quickly raised from below to above the Fermi level \cite{Feve2007}, and a particle is emitted at energy $\Delta/2 > k_{B} T_{el}$ well separated from thermal excitations. For a sine wave,  the rise of the energy level is slower; the electron is then emitted at a lower energy and more prone to antibunch with thermal excitations. This reduces $\delta N_{HBT}$ compared to the square wave. As the transmission $D$ is lowered, the average escape time $\tau$ increases and electron emission
occurs at longer times, corresponding to a higher level of the sine drive. Electrons are then emitted at higher energies and become less sensitive to thermal excitations: $\delta N_{HBT}$ increases as can
be seen by comparing the red and black curves in Fig.2.

\begin{figure}[hhh]
  \centering\includegraphics[width=0.5\textwidth]{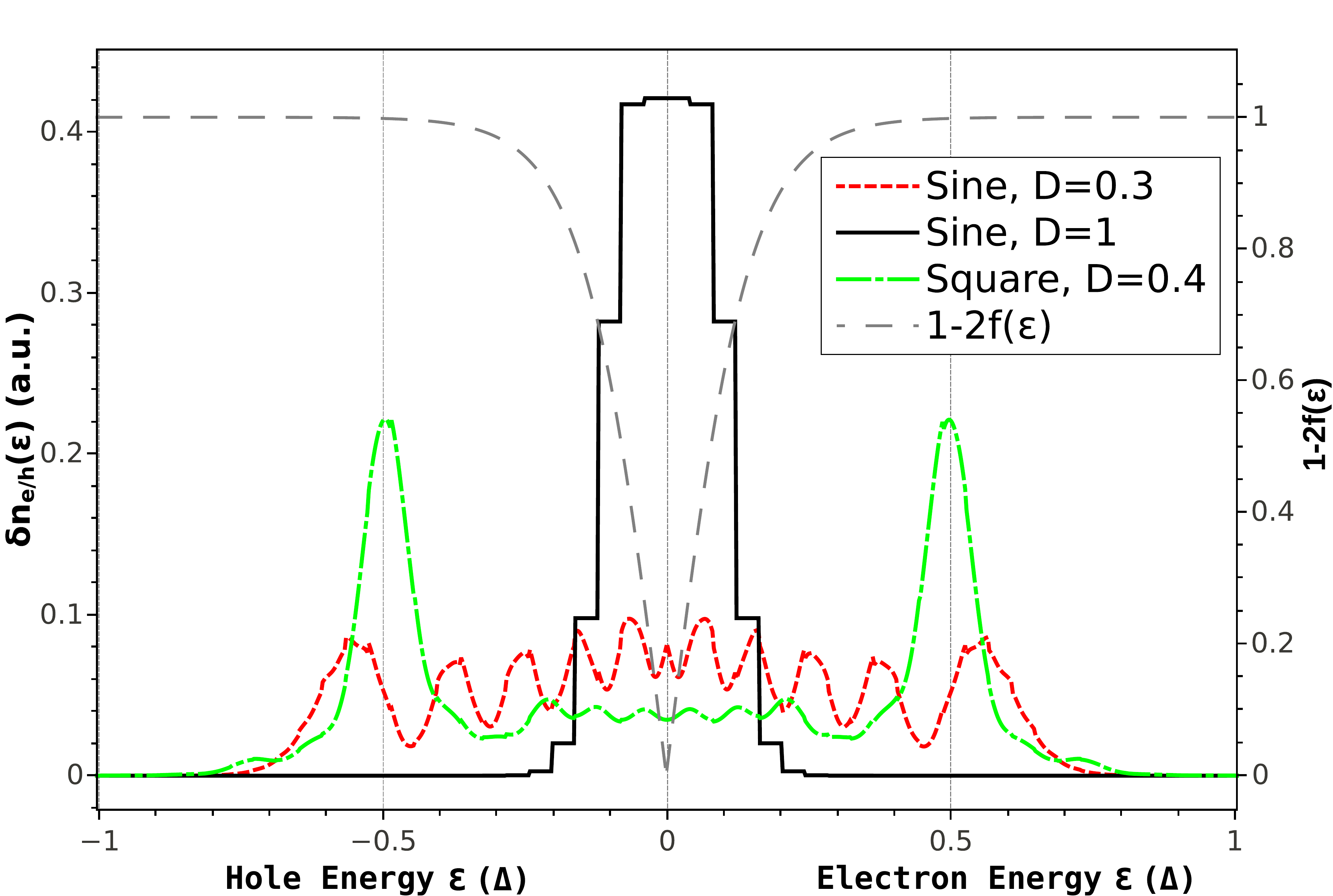}
  \caption{Calculations of $\delta n_{e} (\epsilon)$ (right side) and $\delta n_{h} (\epsilon)$ (left side) at $T_{el,1}=0$ using Floquet scattering theory. The energies are normalized by the dot level spacing $\Delta$. The black line is obtained with a sine drive at $D=1$, the red dashed one with a sine drive at $D=0.3$ and the green dashed one with a square drive at $D=0.4$. Note that in the case of a sine drive at $D=1$, electron/hole pairs generation by the absorption of $n$ photons of energy $h f_d$ is reflected by the steps of width $h f_d$ at energies $0$, $h f_d$, $2h f_d$, $3h f_d$....  At lower transmission $D=0.3$, the plateaus turn into peaks corresponding to the successive attempts of electrons/holes to leave the dot with an attempt frequency of $\Delta/h$. For a square drive, the spectral weight is centered around $\Delta/2$ with a width $\delta \epsilon$ related to transmission $\delta \epsilon \approx \frac{ D \Delta}{2 \pi}$. Note that $\delta n_{e} (\epsilon) = \delta n_{h} (\epsilon)$ (electron-hole symmetry) because the highest energy level of the dot is swept symmetrically around the Fermi energy. The grey dashed line represents $ 1-2 f(\epsilon)  = \tanh{(\frac{\epsilon}{2k_B T_{el}})}$, the fraction of excitations that are effectively counted in the HBT contribution $\delta N_{HBT}$ at temperature $T_{el} = 150$ mK.  }
  \label{Fig3}
\end{figure}

These results can be quantitatively understood using Floquet scattering theory \cite{Floquet} which does not take into account decoherence \cite{Roulleau2008} and relaxation \cite{leSueur2010} which could occur in the $2.5$ $\mu$m propagation length towards the beam-splitter. The energy density $\delta n_{e/h}(\epsilon)$ of the excitations produced by the source can be calculated as a function of the source parameters such as the drive shape or the transmission $D$. Figure 3 presents $\delta n_{e/h}(\epsilon)$ for $T_{el,1}=0$ in arm $1$. These calculations reproduce the previous qualitative discussion. For a square drive, electrons and holes are emitted at energy $\Delta/2$, with an energy width $\delta \epsilon = \frac{D \Delta}{2 \pi}$. Very few excitations are emitted at low energy. For the sine, some excitations are systematically emitted at low energy, especially at high transmission $D=1$.
\begin{figure}[hhh]
  \centering\includegraphics[width=0.5\textwidth]{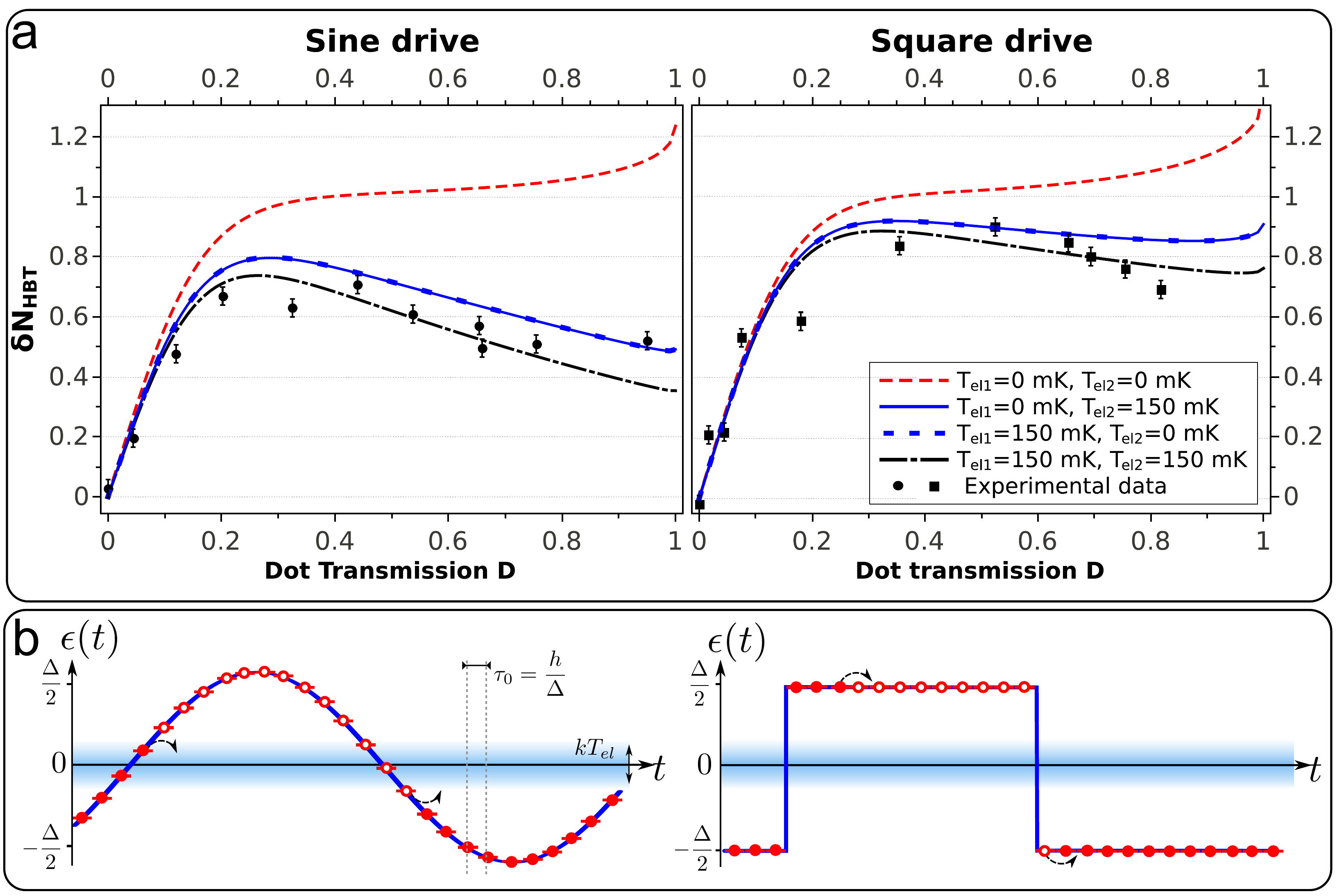}
  \caption{ \textbf{(a)} HBT contribution $\delta N_{HBT}$ as a function of the dot transmission $D$ for a sine (left panel) and a square drive (right panel). The experimental points are represented by circles (sine drive) and squares (square drive). Dashed lines represent numerical evaluations of Eq. (6) using Floquet scattering theory at $T_{el,1}=T_{el,2} =0$ (red dashed line), $T_{el,2}=150$ mK, $T_{el,1} =0$ (blue line), $T_{el,1}=150$ mK, $T_{el,2} =0$ (blue dashed line) , $T_{el,1}=150$ mK, $T_{el,2} =150$ mK (black dashed line).
 \textbf{(b)} Schematic representation of particle emission. On these temporal traces, the filled (respectively empty) red symbols correspond to an occupied (respectively empty) level of the dot.
An electron (respectively a hole) can be emitted every $\tau_0 = h/\Delta$, time to perform one revolution inside the dot, when the occupied (respectively empty) level crosses the Fermi energy ($\epsilon_F=0$). Arrows represent a realization of particle emission. In the case of a sine drive (left hand side), the energy rises slowly with time and particles can be emitted at low energy as compared to the square drive (right hand side) for which
the energy rises abruptly and particles are emitted at energy $\Delta/2$.   }
 \label{Fig4}
\end{figure}

These differences in energy distribution are revealed by HBT interferometry as can be seen on Figure 4, which represents measurements of $\delta N_{HBT}$ as a function of dot transmission $D$ for the square and sine drives. Floquet calculations for $T_{el,1}=T_{el,2}=0$ are presented (red dashed lines), which are almost identical for square and sine, reaching $1$ in the range $0.2 \leq D \leq 0.7$ as expected. The effect of finite temperature in arm $2$, $T_{el,2}=150$ mK is shown by the blue lines, where $T_{el,1} = 0$. As already discussed,  due to thermal excitations, $\delta N_{HBT}$ is lowered. The effect is moderate for the square and important for the sine and decreases by lowering the dot transmission. Blue dashed curves show the effect of temperature in arm $1$ ($T_{el,1}=150$ mK, $T_{el,2}=0$). Remarkably, the role of temperature is identical for both arms : source excitations overlapping with thermal excitations, either in arm $1$ or $2$, are lost. When the actual temperature (extracted from equilibrium noise thermometry of the sample) is introduced on both arms ($T_{el,1}=T_{el,2}=150$ mK, black dashed curves), a good agreement is obtained with the experimental points (symbols) without any adjustable parameters.

As a conclusion, we have realized an HBT partitioning experiment with single electrons. We have used it to count the number of electron and hole excitations emitted per period. Antibunching of low energy excitations with thermal ones is observed which is used to probe the energy distribution of emitted particles.
Since the demonstration of on demand generation of single electron states \cite{Feve2007}, many experiments relying on the coherent manipulation of single to few particles have been suggested \cite{Ol'khovskaya08,  Splettstoesser2009, Grenier2011}. This experiment
is the first realization of an electron optics experiment at the single charge level which
will kick-off the emerging field of electron quantum optics. Furthermore, the HBT geometry benefits from its high versatility. By applying a combination of  $AC+DC$ voltages on input $2$, one can perform a complete tomography \cite{Grenier2011} of the electronic state in input $1$. In particular, one can obtain the detailed spectroscopy of  single excitations which might be affected by interactions during propagation. The electronic variant of the Hong Ou Mandel experiment \cite{Ol'khovskaya08} can also be realized by synchronizing the emission of one electron on each arm. This can be envisioned in the near future thanks to the present experimental realization.

We acknowledge discussions with T. Jonckheere, J. Rech, G. Haack and T. Kontos whom we also thank for a critical reading of the manuscript. This work is supported by the ANR grant '1shot',
ANR-2010-BLANC-0412.

\end{document}